\documentclass{article}

\usepackage{arxiv}

\usepackage[utf8]{inputenc} 
\usepackage[T1]{fontenc}    
\usepackage{hyperref}       
\usepackage{url}            
\usepackage{booktabs}       
\usepackage{amsfonts}       
\usepackage{nicefrac}       
\usepackage{microtype}      
\usepackage{lipsum}
\usepackage{graphicx}
\graphicspath{ {./images/} }
\usepackage{color}

\title{Impact on Public Health Decision Making by Utilizing Big Data Without Domain Knowledge}

\author{
 Miao Zhang \\
  Tandon School of Engineering\\
  New York University\\
  \texttt{miaozhng@nyu.edu} \\
   \And
 Salman Rahman \\
  Tandon School of Engineering\\
  New York University\\
  \texttt{salman@nyu.edu } \\
  \And
 Vishwali Mhasawade \\
  Tandon School of Engineering\\
  New York University\\
  \texttt{vishwalim@nyu.edu } \\
  \And
 Rumi Chunara \\
  School of Global Public Health, Tandon School of Engineering\\
  New York University\\
  \texttt{rumi.chunara@nyu.edu } \\
}

\date{}
\begin{document}

\maketitle
\begin{abstract}
New data sources, and artificial intelligence (AI) methods to extract information from them are becoming plentiful, and relevant to decision making in many societal applications. An important example is street view imagery, available in over 100 countries, and considered for applications such as assessing built environment aspects in relation to community health outcomes. Relevant to such uses, important examples of bias in the use of AI are evident when decision-making based on data fails to account for the robustness of the data, or predictions are based on spurious correlations. To study this risk, we utilize 2.02 million GSV images along with health, demographic, and socioeconomic data from New York City. Initially, we demonstrate that built environment characteristics inferred from GSV labels at the intra-city level may exhibit inadequate alignment with the ground truth. We also find that the average individual-level behavior of physical inactivity significantly mediates the impact of built environment features by census tract, as measured through GSV. Finally, using a causal framework which accounts for these mediators of environmental impacts on health, we find that altering 10\% of samples in the two lowest tertiles would result in a 4.17 (95\% CI 3.84 to 4.55) or 17.2 (95\% CI 14.4 to 21.3) times bigger decrease on the prevalence of obesity or diabetes, than the same proportional intervention on the number of crosswalks by census tract. This work illustrates important issues of  robustness and model specification for informing effective allocation of interventions using new data sources.
\end{abstract}


\section{Introduction}
The proliferation of digital data, alongside artificial intelligence (AI) and machine learning methods to extract information from them, has potential to inform decision-making which can affect large communities and populations in fields such as public health and urban planning. A growing literature has leveraged object detection via deep learning along with image data such as from Google Street View (GSV) to audit neighborhood properties as well as link them to health outcomes. Environmental and urban development features from GSV data, such as types of vegetation, building structures, and road networks, have been linked to health outcomes, including chronic diseases (obesity, diabetes), mental distress, and COVID-19 prevalence~\cite{nguyen2020using, keralis2020health}. At the same time, challenges associated with AI-based predictive models have surfaced. These challenges are particularly evident when dealing with non-representative and biased data~\cite{buolamwini2018gender}. Additionally, there can be statistical challenges, such as making predictions based on spurious correlations~\cite{degrave2021ai}. These challenges are amplified when measuring the effects of environmental attributes (such as from GSV data) on health outcomes, where there can be several intermediate factors interceding the relationship between these exposures and health outcomes. 

In this brief report we study the association between built environment features derived from GSV and mean obesity and diabetes census tract prevalence in New York City (NYC). We find that 
physical inactivity significantly mediates the relationship, and use this causal model to compare the impact of place based interventions on environmental-level attribute versus on average individual behavior by census tract. This work illustrates how using new data sources in concert with public health domain knowledge is essential in order to maintain reliability of models that can inform effective interventions.

\section{Materials and methods}

\paragraph{Data} Using the Google Application Programming Interface (API), GSV images were collected along all streets in New York City (NYC), sampling at points 20m apart\footnote{ https://developers.google.com/maps/documentation/streetview/overview}. At each sampling point, images from the four cardinal directions (North, East, South, West) were collected to form a panorama of the built environment, following previous studies~\cite{kim2021decoding}. In total, 2.02 million images were collected. Street view features (built environment features that may be present in the image): sidewalk and crosswalk, were extracted for all collected images using the Google Vision API\footnote{https://cloud.google.com/vision/docs/labels}. Each image is labeled with sidewalk or crosswalk presence if the API returns a probability score bigger than 0.8 for the label ``sidewalk'' or the label ``zebra crossing'' (label names defined by the API); otherwise no label is applied. The GSV-estimated feature prevalence is computed by taking the ratio of labeled sampling points to all sampling points within each census tract. Sidewalk ground truth data distribution is obtained from NYC OpenData from 2023\footnote{https://data.cityofnewyork.us/City-Government/Sidewalk/vfx9-tbb6}, in which each sidewalk is recorded as a polygon. We compute the prevalence by dividing the area of all sidewalk polygons of a census tract by the area of the census tract, referred to as ground truth sidewalk prevalence. 

Health outcome and health-related behavior data were obtained from the Centers for Disease Control and Prevention PLACES initiative~\cite{CDC_PLACES_2022} at the census tract level ($n=1970$). We selected the health outcome measures of ``Obesity among adults aged $\ge$ 18 years'' (denoted as obesity) and ``Diabetes among adults aged $\ge$ 18 years'' (denoted as diabetes). The health-related behavior ``No leisure-time physical activity among adults aged $\ge$ 18 years'' (denoted as physical inactivity) is tested as mediator. The same demographic and socioeconomic controlling factors from previous work~\cite{nguyen2021leveraging} were used, including socioeconomic status; percentage female, white, and Hispanic; percentage $<$18 and $\ge$65, and median age year. 

The choice of built environment features and health outcomes is motivated by the relevance of sidewalks and crosswalks to obesity and diabetes~\cite{wei2021neighborhood}, as in previous analyses on GSV data~\cite{nguyen2021leveraging, nguyen2022google}. Physical inactivity is used as the mediator, as it reflects lifestyle habits positively related to risk of the studied health outcomes~\cite{jakicic2011obesity, lee2014leisure}. Moreover, physical inactivity has been shown to mediate the effect of built environment features like walkability~\cite{van2016neighborhood}, land-use mix~\cite{xiao2022exploring}, and residential blue space~\cite{pasanen2019neighbourhood}, on health outcomes such as obesity.


\begin{figure}[!ht]
\centering
\includegraphics[width=1\linewidth]{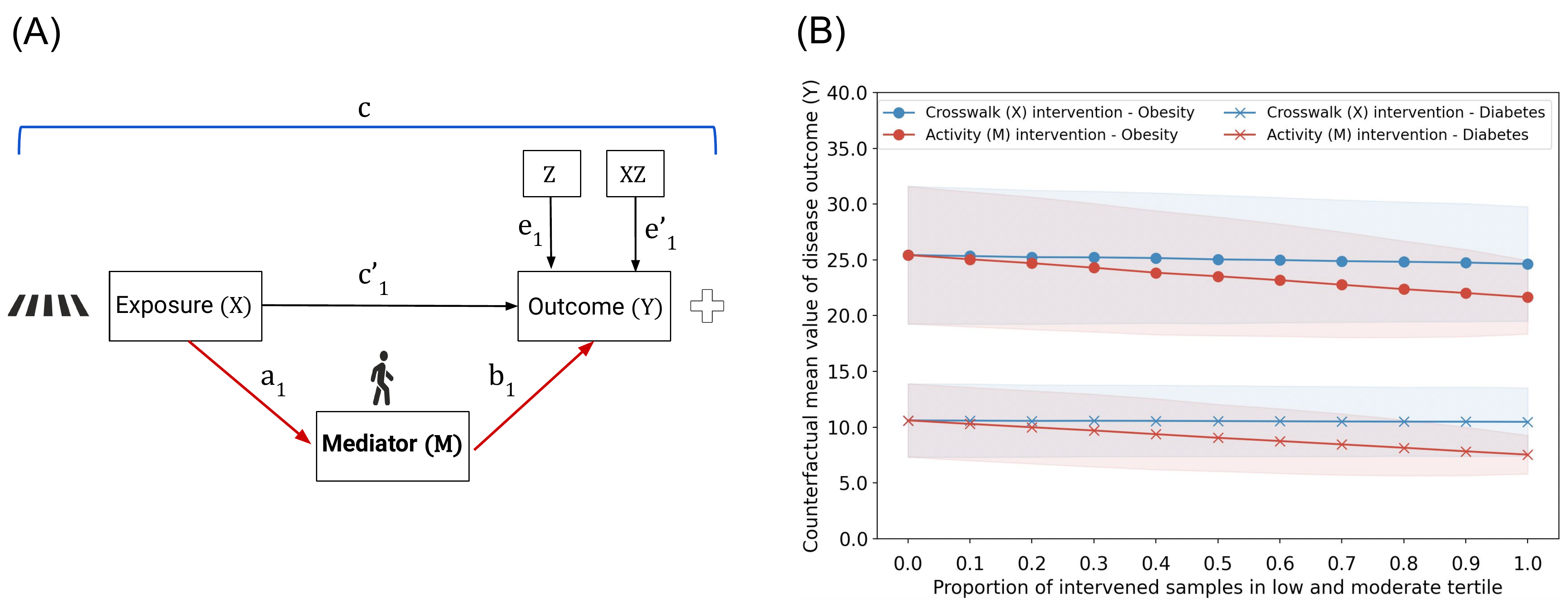}
\caption{\textbf{(A)} Path diagram of the mediation model. The total effect, c, of crosswalk exposure (X) on the outcome (Y), are decomposed into an indirect effect via physical inactivity mediator (M), quantified by a1b1, and direct effect quantified by c’1. \textbf{(B)} Performing health outcome improvement intervention: X and M intervenable variables are grouped into 3 tertiles; census tracts in the low and moderate tertiles are altered to have the same mean as that in the high tertile. The counterfactual Y changes after intervening on X or M, and the outcome mean and $\pm$ standard deviation $\sigma$ is plotted: 1 unit intervention on M (setting 10\% of samples in the two lowest tertiles to have the same mean value of the highest) would result in 4.17 (95\% CI: 3.84, 4.55) times bigger decrease on obesity and 17.2 (95\% CI: 14.4, 21.3) times bigger decrease on diabetes, than the same unit intervention on X. Accounting for the mediation mechanism allows comparative assessment of intervention efficacy, showing that a proportional intervention on physical inactivity has a larger effect on the outcome than the same proportional intervention on the built environment exposure.}
\label{fig:outcome}
\end{figure}

\paragraph{Mediation model and effect analysis} 
A mediation model framework (Fig.~\ref{fig:outcome}A) allows for investigation of different pathways of action of the built environment on health outcomes. Specifically, the mediation framework allows for decoupling the total effect of $X$ on $Y$ (represented by $c$) and the indirect pathway from $X$ to $Y$ through $M$. The mediation effect of $M$ is captured by $a_1b_1$ which measures how much the relationship between sidewalk and crosswalk prevalence) and health outcome rates (obesity and diabetes) are delivered via individual behavior (physical inactivity). Linear models were used to examine the association between $X$ and $Y$, conditioning on covariates ($Z$; demographic and socioeconomic characteristics).

\paragraph{Counterfactual outcome} To investigate the effect of disregarding mediation of environmental attributes measured via GSV on decision making, we investigated the change in outcome (mean value of obesity and diabetes) via intervention at the i) environment level, and ii) physical activity level. For the built environment intervention, we split samples into 3 tertiles: high, moderate, and low based on the distribution of $X$, and define 1 unit of intervention as setting 10\% of samples in low and moderate tertile to have the same mean value of $X$ as the high tertile. The intervention on individual-level behavior is performed with the same procedure on the $M$ distribution. The counterfactual $Y$ after intervention is computed using coefficients of the established mediation model (SI Appendix). 

\paragraph{GSV feature representativeness} We analyzed how well the GSV-estimated sidewalk prevalence corresponds to ground truth sidewalk prevalence by computing the Pearson correlation coefficient between the two at NYC and borough levels. We performed qualitative examination by selecting census tracts with (top 5\%) largest discrepancies between the two measures and manually examined errors. Errors were summarized via common false positives and negatives in GSV-labelled images.

\begin{table*}[tbhp]
\setlength{\tabcolsep}{4pt}
\centering
\small
\scalebox{0.96}{
\begin{tabular}{lrrrr}
Health outcome: Obesity \\
\midrule
& \textcolor{blue}{Total effect ($c$)} & \textcolor{red}{Mediation effect ($a_1b_1$)} & Effect on mediator ($a_1$) & Mediator effect on outcome ($b_1$) \\
\midrule
Crosswalk & -1.11 ***  & -2.46 *** & -0.682 *** & 3.60 *** \\
& [-1.36, -0.860] & [-3.58, -1.50] & [-0.896, -0.468] & [3.21, 4.00] \\
Sidewalk & 0.00125 & -0.996 ** & -0.253 ** & 3.93 *** \\
& [-0.187, 0.189] & [-1.82, -0.305] & [-0.420, -0.0858] & [3.55, 4.33] \\
\midrule
Health outcome: Diabetes \\
\midrule
& \textcolor{blue}{Total effect ($c$)} & \textcolor{red}{Mediation effect ($a_1b_1$)} & Effect on mediator ($a_1$) & Mediator effect on outcome ($b_1$) \\
\midrule
Crosswalk & -0.153 *** & -1.96 *** & -0.682 *** & 2.88 *** \\
& [-0.243, -0.0626] & [-2.67, -1.30] & [-0.896, -0.468] & [2.77, 2.98] \\
Sidewalk & 0.0246 & -0.721 ** & -0.253 ** & 2.85 *** \\
& [-0.0489, 0.0981] & [-0.236, -1.24] & [-0.420, -0.0858] & [2.75, 2.95] \\
\bottomrule
\end{tabular}
}
\caption{Effect sizes and 95\% CIs (in brackets) for the model that tests the total effect of built environment exposure ($X$): Crosswalk and Sidewalk on health outcome ($Y$): Obesity and Diabetes, and the models that test the mediation effect of inctivity ($M$) in the relationship of exposure with outcome.  *** $p < 0.001$, ** $p < 0.01$, * $p < 0.05$.}
\label{table1}
\end{table*}

\section{Results}

\textbf{GSV-derived features association with health outcomes.} The total effect of the crosswalk feature, as measured by GSV, is negative, indicating that a higher crosswalk density is associated with lower disease prevalence. Further, the effect on health outcomes is significant, with a bigger effect on obesity ($c=-1.11$) than diabetes ($c=-0.153$). These findings are consistent with previous studies based on GSV-estimated crosswalk feature~\cite{keralis2020health, nguyen2021leveraging, nguyen2022google}. However, no significant associations were found between GSV-estimated sidewalk feature and health outcomes (Table~\ref{table1}), in contrast to a previous study which analyzed the association nation-wide ~\cite{yue2022using}.

\noindent \textbf{Physical inactivity is a significant mediator.} The direction of all mediation effects ($a_1b_1$) are negative, operating through a negative exposure effect on the mediator ($a_1$) and positive effect of the mediator on the outcomes ($b_1$) (Table~\ref{table1}). The total effect of GSV-measured crosswalk prevalence is significantly mediated by physical inactivity prevalence ($a_1b_1 = -2.46/-1.96$, for obesity/diabetes), and physical inactivity entirely mediates the total effect of sidewalk prevalence on health outcomes ($a_1b_1 = -0.996/-0.721$), as neither simple association between sidewalk prevalence and obesity/diabetes are significant (Table~\ref{table1}). That is, decreased obesity or diabetes prevalence linked to increased crosswalk or sidewalk prevalence by census tract measured through GSV, can be largely accounted for by whether individuals in that census tract have increased physical activity. 

\noindent \textbf{Intervening on physical inactivity versus street view features.} Analysis shows that a 1 unit decrease of physical inactivity would result in a 4.17 (95\% CI: 3.84, 4.55) times bigger decrease of obesity prevalence and 17.2 (95\% CI: 14.4, 21.3) times bigger decrease of diabetes prevalence, compared to the same intervention on crosswalk prevalence (Fig.~\ref{fig:outcome} (B)).

\begin{figure}[!ht]
\centering
\includegraphics[width=0.7\linewidth]{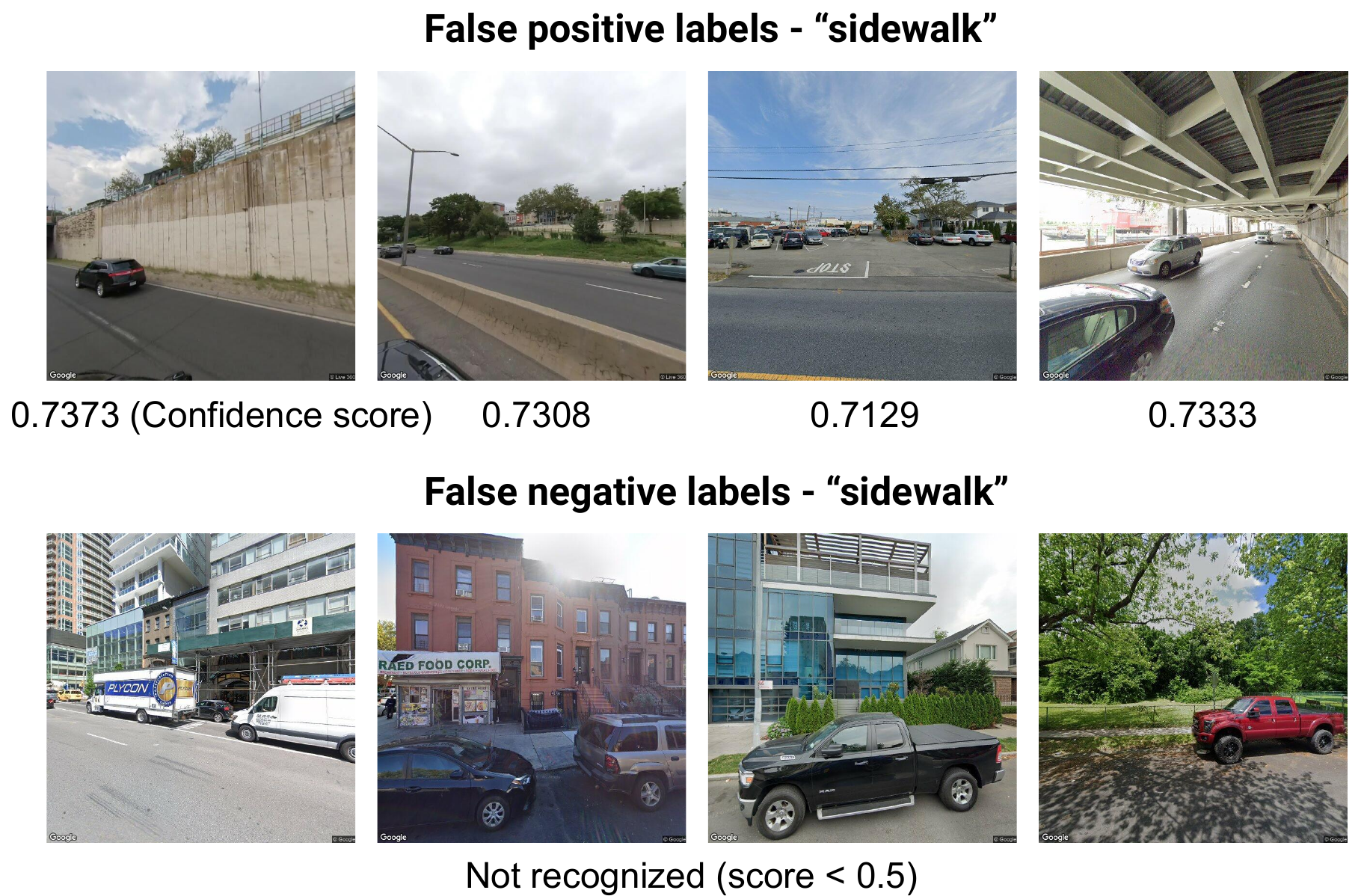}
\caption{Common false positive and negative errors for GSV labels for sidewalk. Top row: GSV incorrectly returns high confidence of a sidewalk, as side lanes of highways, bridges, or parking lots have similar shape to and are easily confused with sidewalks. Bottom row GSV labels are incorrectly low confidence in recognizing sidewalk because of the obstruction from vehicles, constructions, or tree shade. Data source: Google Street View images.}
\label{fig:robustness}
\end{figure}

\noindent \textbf{Street view features may not represent the built environment.} 
Pearson correlation coefficient between GSV-estimated and ground truth sidewalk prevalence was 0.214 ($p<0.001$) at the city level. At the borough level, the correlation is significant ($p<0.05$) with coefficient 0.361 for Bronx, 0.386 for Manhattan, 0.327 for Queens, 0.331 for Staten Island, but not significant for Brooklyn. From qualitative examination we found that GSV may falsely report sidewalks near highways, expressways, or bridges, or absence in places with obstruction on sidewalks (Fig.~\ref{fig:robustness}). 


\section{Discussion}
Growing amounts of digital data can be useful to inform decision making, but our results show the key finding that simply using associations as suggested in previous work, can misappropriate the utility of the information. Specifically, we showed that mean physical inactivity by census tract significantly mediates the impact of built environment features measured through GSV, on prevalence of obesity and diabetes. Further, we show that accounting for the mediator can be important in order to improve the target and efficacy of interventions.  

Our study makes several important advances. Importantly, we examine GSV coverage compared to ground truth at a city level, opposed to existing studies which compare areas by qualitative review. In doing so we show that at a granular level, built environment features based on labels from Google Street view may not match with ground truth at this spatial granularity. 
We also utilize a causal framework to show how conclusions (e.g. informing interventions) based on the data must take such mechanisms into account; especially relevant to those at multiple levels (built environment versus person-level).

In conclusion, this study highlights the potential of digital data sources like GSV in enhancing public health research, while also pointing out the need for careful consideration of data limitations and the complex dynamics between the built environment, individual behavior, and health outcomes. Future research should focus on addressing these challenges and exploring innovative strategies to leverage digital data for more effective public health interventions.

\bibliographystyle{unsrt}  

\bibliography{references}

\begin{thebibliography}{10}

\bibitem{nguyen2020using}
Quynh~C Nguyen, Yuru Huang, Abhinav Kumar, Haoshu Duan, Jessica~M Keralis, Pallavi Dwivedi, Hsien-Wen Meng, Kimberly~D Brunisholz, Jonathan Jay, Mehran Javanmardi, et~al.
\newblock Using 164 million google street view images to derive built environment predictors of covid-19 cases.
\newblock {\em International journal of environmental research and public health}, 17(17):6359, 2020.

\bibitem{keralis2020health}
Jessica~M Keralis, Mehran Javanmardi, Sahil Khanna, Pallavi Dwivedi, Dina Huang, Tolga Tasdizen, and Quynh~C Nguyen.
\newblock Health and the built environment in united states cities: Measuring associations using google street view-derived indicators of the built environment.
\newblock {\em BMC public health}, 20(1):1--10, 2020.

\bibitem{buolamwini2018gender}
Joy Buolamwini and Timnit Gebru.
\newblock Gender shades: Intersectional accuracy disparities in commercial gender classification.
\newblock In {\em Conference on fairness, accountability and transparency}, pages 77--91. PMLR, 2018.

\bibitem{degrave2021ai}
Alex~J DeGrave, Joseph~D Janizek, and Su-In Lee.
\newblock Ai for radiographic covid-19 detection selects shortcuts over signal.
\newblock {\em Nature Machine Intelligence}, 3(7):610--619, 2021.

\bibitem{kim2021decoding}
Jae~Hong Kim, Sugie Lee, John~R Hipp, and Donghwan Ki.
\newblock Decoding urban landscapes: Google street view and measurement sensitivity.
\newblock {\em Computers, Environment and Urban Systems}, 88:101626, 2021.

\bibitem{CDC_PLACES_2022}
Centers for Disease~Control, National Center for Chronic Disease~Prevention Prevention, and Division of Population~Health Health~Promotion.
\newblock Places: Local data for better health, zcta data 2022, 2022.

\bibitem{nguyen2021leveraging}
Quynh~C Nguyen, Jessica~M Keralis, Pallavi Dwivedi, Amanda~E Ng, Mehran Javanmardi, Sahil Khanna, Yuru Huang, Kimberly~D Brunisholz, Abhinav Kumar, and Tolga Tasdizen.
\newblock Leveraging 31 million google street view images to characterize built environments and examine county health outcomes.
\newblock {\em Public Health Reports}, 136(2):201--211, 2021.

\bibitem{wei2021neighborhood}
Junxiang Wei, Yang Wu, Jinge Zheng, Peng Nie, Peng Jia, and Youfa Wang.
\newblock Neighborhood sidewalk access and childhood obesity.
\newblock {\em Obesity reviews}, 22:e13057, 2021.

\bibitem{nguyen2022google}
Quynh~C Nguyen, Tom Belnap, Pallavi Dwivedi, Amir Hossein~Nazem Deligani, Abhinav Kumar, Dapeng Li, Ross Whitaker, Jessica Keralis, Heran Mane, Xiaohe Yue, et~al.
\newblock Google street view images as predictors of patient health outcomes, 2017--2019.
\newblock {\em Big data and cognitive computing}, 6(1):15, 2022.

\bibitem{jakicic2011obesity}
John~M Jakicic and Kelliann~K Davis.
\newblock Obesity and physical activity.
\newblock {\em Psychiatric Clinics}, 34(4):829--840, 2011.

\bibitem{lee2014leisure}
Duck-Chul Lee, Russell~R Pate, Carl~J Lavie, Xuemei Sui, Timothy~S Church, and Steven~N Blair.
\newblock Leisure-time running reduces all-cause and cardiovascular mortality risk.
\newblock {\em Journal of the American College of Cardiology}, 64(5):472--481, 2014.

\bibitem{van2016neighborhood}
Jelle Van~Cauwenberg, Veerle Van~Holle, Ilse De~Bourdeaudhuij, Delfien Van~Dyck, and Benedicte Deforche.
\newblock Neighborhood walkability and health outcomes among older adults: The mediating role of physical activity.
\newblock {\em Health \& place}, 37:16--25, 2016.

\bibitem{xiao2022exploring}
Yang Xiao, Sijia Chen, Siyu Miao, and Yifan Yu.
\newblock Exploring the mediating effect of physical activities on built environment and obesity for elderly people: Evidence from shanghai, china.
\newblock {\em Frontiers in Public Health}, 10:853292, 2022.

\bibitem{pasanen2019neighbourhood}
Tytti~P Pasanen, Mathew~P White, Benedict~W Wheeler, Joanne~K Garrett, and Lewis~R Elliott.
\newblock Neighbourhood blue space, health and wellbeing: The mediating role of different types of physical activity.
\newblock {\em Environment international}, 131:105016, 2019.

\bibitem{yue2022using}
Xiaohe Yue, Anne Antonietti, Mitra Alirezaei, Tolga Tasdizen, Dapeng Li, Leah Nguyen, Heran Mane, Abby Sun, Ming Hu, Ross~T Whitaker, et~al.
\newblock Using convolutional neural networks to derive neighborhood built environments from google street view images and examine their associations with health outcomes.
\newblock {\em International Journal of Environmental Research and Public Health}, 19(19):12095, 2022.

\end{thebibliography}






\end{document}